\documentclass[prb,aps,twocolumn,showpacs,preprintnumbers,amsmath,amssymb, superscriptaddress]{revtex4-2}

\usepackage[makeroom]{cancel}

\usepackage{amsmath}    
\usepackage{amssymb}
\usepackage{graphicx}   
\usepackage{verbatim}   
\usepackage{color}      
\usepackage{hyperref}   
\usepackage[normalem]{ulem}
\usepackage{natbib}
\usepackage{enumitem}
\usepackage{empheq}

\hypersetup{colorlinks,linkcolor=blue,urlcolor=blue,citecolor=blue}

\definecolor{darkgreen}{rgb}{0,0.5,0}
\definecolor{purple}{rgb}{0.35,0,0.35}
\definecolor{orange}{rgb}{1,0.5,0}
\definecolor{darkred}{rgb}{.7,0,0}
\definecolor{darkblue}{rgb}{0,0,.3}
\definecolor{grey}{rgb}{.6,.6,.6}
\definecolor{dimgreen}{rgb}{0.2,0.6,0.1}

\definecolor{darkgreen}{rgb}{0,0.5,0}

\begin{document}

\title{Quantum quench dynamics in the Luttinger liquid phase of the Hatano-Nelson model}

\author{Bal\'azs D\'ora}
\email{dora.balazs@ttk.bme.hu}
\affiliation{MTA-BME Lend\"ulet Topology and Correlation Research Group,
Budapest University of Technology and Economics, M\H uegyetem rkp. 3., H-1111 Budapest, Hungary}
\affiliation{Department of Theoretical Physics, Institute of Physics, Budapest University of Technology and Economics, M\H uegyetem rkp. 3., H-1111
Budapest, Hungary}
\author{Mikl\'os Antal Werner}
\affiliation{Department of Theoretical Physics, Institute of Physics, Budapest University of Technology and Economics, M\H uegyetem rkp. 3., H-1111
Budapest, Hungary}
\affiliation{Strongly Correlated Systems 'Lend\" ulet' Research Group, Wigner Research Centre for Physics, P.O. Box 49, 1525 Budapest, Hungary}
\affiliation{MTA-BME Quantum Dynamics and Correlations Research Group, Institute of Physics, Budapest University of Technology and Economics, M\H uegyetem rkp. 3., H-1111, Budapest, Hungary}
\author{C\u{a}t\u{a}lin Pa\c{s}cu Moca}
\affiliation{MTA-BME Quantum Dynamics and Correlations Research Group, Institute of Physics, Budapest University of Technology and Economics, M\H uegyetem rkp. 3., H-1111, Budapest, Hungary}
\affiliation{Department  of  Physics,  University  of  Oradea,  410087,  Oradea,  Romania}

\date{\today}

\begin{abstract}

    We investigate the quantum quench dynamics of the interacting Hatano-Nelson model with open boundary conditions using both abelian bosonization and 
numerical methods. Specifically, we follow the evolution of the particle density and current profile in real space over time by turning the imaginary 
vector potential on or off in the presence of weak interactions.
Our results reveal spatio-temporal Friedel oscillations in the system with light cones propagating ballistically from the 
open ends, accompanied by local currents of equal magnitude for both switch off and on protocols. Remarkably, the bosonization method accurately accounts 
for the density and current patterns with a single overall fitting parameter. The continuity equation is satisfied by the long wavelength part of the 
density and current, despite the non-unitary time evolution when the Hatano-Nelson term is switched on.
\end{abstract}

\maketitle

\section{Introduction}

The non-Hermitian phenomena have been gaining significant attention in recent years, mainly due to their ability to exhibit unexpected features and their 
applicability to a broad range of classical and quantum systems~\cite{ashidareview,Bergholtz2021}. These features include exceptional points~\cite{heiss,hodaei,ding2022}, 
which refer to the points in the parameter space where two or more eigenvalues and eigenvectors of the matrix Hamiltonian coalesce. At these points, the 
behavior of the system can change drastically. PT-symmetry breaking~\cite{ElGanainy2018} is another important feature of non-Hermitian systems, leading to 
the eigenvalues and eigenvectors becoming complex and the system becoming unstable, resulting in phenomena such as unidirectional invisibility, non-reciprocal 
energy transfer, and enhanced sensitivity. Additionally, non-Hermitian systems exhibit non-trivial topological phenomena, leading to the emergence of edge modes~\cite{rotter,gao2015,zhou18,zeuner,gongprx,lee2016,takasu,fruchart}.

The Hatano-Nelson model is one of the earliest models in non-Hermitian physics~\cite{hatanonelson2, hatanonelson1}. It features non-interacting particles on a quantum ring, subject to an imaginary vector potential that renders the problem non-Hermitian. Initially, the focus was on persistent current and localization within this non-Hermitian context. 
However, this field has since experienced rapid expansion, with numerous studies being conducted on the non-Hermitian skin effect and related phenomena in the Hatano-Nelson model, as well as in other non-Hermitian systems such as photonic crystals~\cite{Feng2017, Ozawa.2019} and electronic systems~\cite{Nagai.2020}.

The non-Hermitian skin effect is characterized by the unusual localization of all eigenstates, as each single particle eigenstate becomes exponentially localized at the boundaries of 
the system, even without the presence of disorder~\cite{lee2016, yao2018, kunst2018}. While this effect is primarily observed at the single particle level, many studies have explored 
its behavior in a many-body context using analytical Bethe ansatz~\cite{fukui, mao2022}, bosonization~\cite{dorahn}, and numerical methods~\cite{zhang2022, alsallom, lee2020, 
hamazaki, mu2020, zhangprb2020, wang2022}. However, the dynamics of the model, particularly in the interacting case, have received less attention. 
In Ref. \onlinecite{kawabata} 
the evolution of the entanglement entropy and the transition from a volume to an area law in the non-interacting Hatano-Nelson model was studied during quantum quench dynamics, but 
fewer studies have investigated the dynamical properties of the interacting case.

Our motivation to investigate the quench dynamics of the interacting Hatano-Nelson model stems from the need to understand its dynamical properties, such as the propagation of the light cone and the spatio-temporal density and current profiles when the imaginary vector potential is turned on or off. 

In general, in a quench problem, the initial state of the system is prepared in the ground state of the Hamiltonian with certain parameters, and then suddenly the parameters of the Hamiltonian are changed. This sudden change drives the system out of equilibrium, and the system's dynamics are governed by the new Hamiltonian.
In general, non-Hermitian Hamiltonians may not have a well-defined ground state in the traditional sense because they do not guarantee real eigenvalues or orthogonal eigenvectors. 
Instead, they often exhibit complex eigenvalues and non-orthogonal eigenvectors. Consequently, the concept of ground state, which relies on the lowest real eigenvalue and its 
corresponding eigenvector, is not directly applicable to non-Hermitian Hamiltonians. However, for some specific non-Hermitian systems such as the  Hatano-Nelson model, the 
PT-symmetry~\cite{ElGanainy2018,bender2007} guarantees that it has real eigenvalues and possess a ground state. 

To achieve this, we will employ bosonization, a powerful technique used to study low-dimensional systems, including one-dimensional systems of interacting fermions adapted to the non-Hermitian realm~\cite{dorahn,yamamoto,Affleck2004,Hofstetter2004,yamamoto2}.
Additionally, we will use numerical tools like the density matrix renormalization group\cite{dmrgmps} (DMRG) to obtain the ground state and time-evolving block decimation (TEBD)~\cite{Vidal-2007} for analyzing the system's dynamics.

The structure of the paper is as follows: Section~\ref{sec:bosonization} provides an introduction to the bosonized version of the Hatano-Nelson model, including the construction of the vertex function, which enables the calculation of the spatio-temporal dependence of the average occupations. Sections~\ref{sec:on} and \ref{sec:off} are dedicated to the discussion of two types of quenches, in which the imaginary vector potential is either switched on or off. The relationship with the continuity equation is explored in Section~\ref{sec:ce}, and a comparison between the bosonization and numerical approaches is presented in Section~\ref{sec:numerics}.

\section{Bosonized Hatano-Nelson model} \label{sec:bosonization}

One can construct an effective low-energy Hamiltonian in the presence of an imaginary vector potential~\cite{giamarchi,cazalillaboson,nersesyan,dorahn} using standard abelian bosonization as
\begin{gather}
H=\int_0^L \frac{dx}{2\pi} v\left[K(\pi\Pi(x)-ih)^2+\frac 1K (\partial_x\phi(x))^2\right],
\label{hamboson}
\end{gather}
where $\Pi(x)$ and $\phi(x)$ are the dual fields satisfying 
the regular commutation relation \cite{cazalillaboson}, $[\Pi(x),\phi(x')]=i\delta(x-x')$.
This non-hermitian Hamiltonian is brought to conventional Luttinger liquid (LL) form by applying a similarity transformation~\cite{dorahn}, which eliminates the vector 
potential terms using $S^{-1}HS$ with
\begin{gather} 
S=\exp\left(\frac{h}{\pi}\int_0^L\phi(x')dx'\right),
\label{smatrix}
\end{gather}
and $S^{-1}$ is obtained from  $S$ after $h\rightarrow -h$ change.
The resulting Hamiltonian is diagonalized after introducing canonical bosonic fields~\cite{giamarchi} as 
\begin{gather}
H_b=\sum_{q>0}\omega(q)b^\dagger_qb_q,
\label{hboson}
\end{gather}
 and the long wavelength part of the local charge density is $\partial_x \phi(x)/\pi$
with
\begin{gather}
 \phi(x)=i\sum_{q>0}\sqrt{\frac{\pi K}{qL}}\sin(qx)\left[b_q-b^\dagger_q\right]
\label{phix}
\end{gather}
for open boundary conditions (OBC)\cite{cazalillarmp}
and $K$ the LL parameter\cite{giamarchi}, which carries all the non-perturbative effects of interaction 
and $\omega(q)=vq$ with $v$ the Fermi velocity in the interacting systems and $q=l\pi/L$ with $l=1,2,3\dots$.
The ground state of Hamiltonian~\eqref{hboson} is the bosonic vacuum $|0\rangle$, and the ground state of the original non-hermitian Hamiltonian~\eqref{hamboson} is obtained by applying  $S$  to the vacuum state $|0\rangle$,
\begin{equation}
    |\Phi\rangle=\frac{S|0\rangle}{\sqrt{\langle 0|S^2| 0\rangle}},
    \label{eq:phi_0}
\end{equation}
 where in the denominator we use the hermiticity of $S$ as defined in Eq. \eqref{smatrix}. This indicates that that the low energy 
effective theory of the interacting Hatano Nelson model is a Luttinger liquid with collective bosonic excitations, similarly to hermitian systems~\cite{giamarchi,cazalillaprl,cazalillarmp}.

We investigate the vertex operator~\cite{delft},
given by
\begin{gather}
G_\lambda(x,t)=\left\langle \Phi(t) \left|\exp\left( i\lambda \phi(x)\right)\right |\Phi(t) \right\rangle,
\label{charfunc}
\end{gather}
where $|\Phi(t)\rangle$ is the time evolved wavefunction after the quench as we discuss below.
From this, the long wavelength and $2k_F$ oscillating part of the density are obtained as 
\begin{subequations}
\begin{gather}
n_0(x,t)=\frac{1}{\pi}\lim_{\lambda\rightarrow 0}\partial_x G_\lambda(x,t)/(i\lambda),\label{n0eq}\\
n_{2k_f}(x,t)=G_2(x,t).\label{nn}
\end{gather}
\end{subequations}
For the quench problems that we address, we are able to provide analytical expression for the 
vertex function in Eq.~\eqref{charfunc} and consequently for the time dependent particle densities.

\section{Switching off the non-hermitian term ($h\ne 0 \to h=0$)}\label{sec:off}
In the first configuration we prepare the system in the ground state as defined by Eq.~\eqref{eq:phi_0} with an imaginary vector potential present. The quench consists in turning off the vector potential and allowing the system to evolve unitarily under Hermitian Hamiltonian~\eqref{hamiltontb} with zero imaginary vector potential, $h=0$  \cite{lancaster}. 
We coin this as the 'switch off' protocol.
 The time evolution 
is  therefore 
\begin{gather}
|\Phi(t)\rangle =\frac{\exp(-iH_bt)S|0\rangle}{\sqrt{\langle 0|S^2| 0\rangle}}.
\end{gather}
The vertex operator is evaluated by realizing that for any $A$ and $B$, two operators that are  linear in the bosonic field as 
for example in Eq. \eqref{phix}, the identity
\begin{gather}
\exp(A)\exp(B)\exp(A)=\exp(2A+B)
\label{identity}
\end{gather}
holds. (Its derivation follows from  using the Baker – Campbell – Hausdorff formula~\cite{delft} twice).
Eventually, we get
\begin{gather}
G_\lambda(x,t)=\frac{\left\langle 0 \left|\exp\left( i\lambda \phi(x,t)+\frac{2h}{\pi} \int_0^L\phi(x')dx'\right)\right |0 \right\rangle}{\left\langle 0 
\left|\exp\left(\frac{2h}{\pi} \int_0^L\phi(x')dx'\right)\right |0 \right\rangle}
,
\end{gather}
where $\phi(x,t)=\exp(iH_bt)\phi(x)\exp(-iH_bt)$ is the time dependent bosonic field. 
This amounts to use $b_q(t)=b_q\exp(-i\omega(q)t)$ and $b^\dagger_q(t)=b_q\exp(i\omega(q)t)$
in Eq. \eqref{phix}.

\begin{figure}[t!]
\centering
\includegraphics[width=7cm]{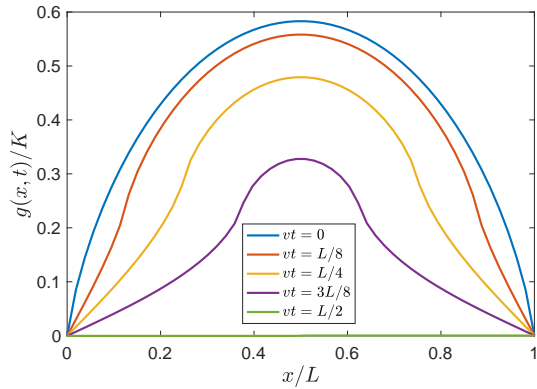}
\caption{The time evolution of $g(x,t)$ from Eq. \eqref{gx} is visualized, carrying all the effects of the Hatano-Nelson term for quarter time period. For 
longer times, it exhibits a sign 
change before reverting to the above pattern.}
\label{gxt}
\end{figure}

Using the standard trick of $\langle\exp(A)\rangle=\exp(\langle A^2\rangle/2)$, which is valid for a Gaussian wavefunction and an operator $A$ being linear in the bosonic field, the expectation value of the vertex operator is evaluated to yield
\begin{gather}
\ln G_\lambda(x,t)=-\frac{\lambda^2}{2}\langle 0|\phi(x,t)^2|0\rangle+\nonumber\\
+\frac{ih\lambda L}{\pi} \int_0^L \frac{dx'}{L}\langle 0|\left\{\phi(x,t),\phi(x')\right\}|0\rangle,
\label{g1}
\end{gather} 
where the expectation values can be easily evaluated~\cite{giamarchi} and $\{A,B\}$ denotes the anticommutator.
Altogether, it is rewritten as
\begin{gather}
\ln G_\lambda(x,t)=-\frac{\lambda^2}{2}C_\phi(x)+\frac{2i \lambda hL}{\pi}g(x,t)
\end{gather}
with
\begin{gather}
C_\phi(x)=\langle 0|\phi(x,t)^2|0\rangle=\frac K2 \ln\left(\frac{2L}{\pi\alpha}\sin\left(\frac{\pi x}{L}\right)\right)
\end{gather}
being time independent since $\exp(-iH_bt)|0\rangle=|0\rangle$ and valid for $\alpha\ll x\ll L$ with  $\alpha$ is the short distance cutoff, 
a remnant of the lattice constant when taking the continuum limit.
 The second term involving the anticommutator gives
\begin{gather}
g(x,t)\equiv \int_0^L \frac{dx'}{2L}\langle 0|\left\{\phi(x,t),\phi(x')\right\}|0\rangle=\nonumber\\
=\frac{K}{2\pi} ~\textmd{Im}\sum_{\beta,\sigma=\pm}\beta ~\textmd{polylog}\left(2, \beta \exp\left(\frac{i\pi (x+\sigma vt)}{L}\right)\right),
\label{gx}
\end{gather}
where polylog$(2,x)$ is the 2nd order polylogarithm\cite{gradstein}. This is valid in the scaling limit, 
when the space-time parameters $x$ and $vt$ and their combinations, including the light cones at $x\pm vt$ differ significantly from $\alpha$ and $L$.
The $g(x,t)$ function is periodic in both $x$ and $vt$ with period $2L$.
The expectation value of the vertex operator, $G_\lambda(x,t)$, is also related to the generating function of the quantity $\phi$. It features the usual contribution
from a LL with open boundary condition, $C_\phi(x)$, and an additional piece coming from the Hatano-Nelson term, $g(x,t)$. As we show below, similar properties
characterize the dual field $\Theta$ as well. In Fig.~\ref{gxt}, we plot the function $g(x,t)$, which captures all effect of non-hermiticity within the validity
of a low energy theory.

\begin{figure}[t!]
\centering
\includegraphics[width=8cm]{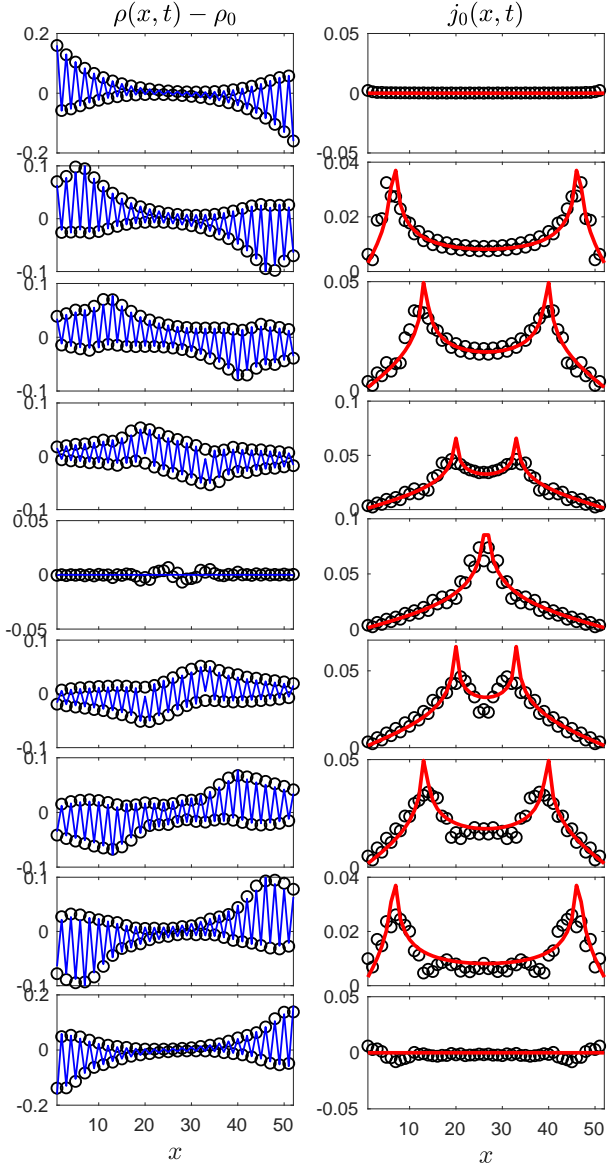}
\caption{Real space density profile (left) and current (right) for the interacting Hatano-Nelson model with $U=0.5J$, 
$L=52$ and times $t=k~L/8v$ with $k=0:1:8$ from top to bottom after switching off  
$h=0.1$. The Friedel oscillations are plotted using $c=0.42$ and $\delta=0$. The circles denote the numerical data from DMRG, the only overall 
fitting parameter is $c$ 
for the oscillating part of the 
particle density, the long wavelength parts contain no fitting parameter, as seen from Eqs. \eqref{n0} and \eqref{j0}.}
\label{friedel}
\end{figure}

The homogeneous part of the particle density exhibits ballistic propagation of the initial non-hermitian parameter induced density profile as
\begin{gather}
n_0(x,t)=-\frac{Kh}{\pi^2}\sum_{\sigma=\pm}\ln\left|\tan\left(\frac{\pi (x+\sigma vt)}{2L}\right)\right|,
\label{n0}
\end{gather}
and is directly proportional to $\partial_x g(x,t)$ through Eq. \eqref{n0eq}. Initially, light cones appear at around the boundaries of the system and start propagating to the other ends with time.
For $vt=L(k+\frac 12)$ with integer $k$, this homogeneous part vanishes identically and we are left only with the $2k_F$ oscillating part of the density, as shown in 
Fig. \ref{friedel}. 
Putting everything together, the total time dependent particle density is
\begin{gather}
\rho(x,t)=\rho_0+n_0(x,t)+\nonumber\\
+c\left(\frac{\pi\alpha}{2L\sin\left(\frac{\pi 
x}{L}\right)}\right)^K\sin\left(2k_Fx-\frac{4hL}{\pi}g(x,t)+\delta\right),
\label{totaln}
\end{gather}
where $\rho_0$ represents the homogeneous background, $c$ is an overall constant, which cannot be obtained from the low energy theory and $\delta$ denotes the phase shift.

The redistribution of charge is accompanied by local currents flowing in the system. Their effect can be captured by evaluating the other vertex operator,
\begin{gather}
F_\lambda(x,t)=\left\langle \Phi(t) \left|\exp\left( i\lambda \Theta(x)\right)\right |\Phi(t) \right\rangle,
\end{gather}
where
\begin{gather}
\Theta(x)=\sum_{q>0}\sqrt{\frac{\pi}{KqL}}\cos(qx)\left[b_q+b^\dagger_q\right].
\label{thetax}
\end{gather}
Within the low energy effective theory, the  $2k_f$ oscillating part of the particle current is usually highly irrelevant (its scaling dimension is large compared to 
its
long wavelength counterpart), therefore we refrain from analyzing it.
This yields the local current\cite{giamarchi} through
\begin{gather}
j_0(x,t)=\frac{vK}{\pi}\lim_{\lambda\rightarrow 0}\partial_x F_\lambda(x,t)/(i\lambda).
\end{gather}

Following similar steps, we obtain
\begin{gather}
\ln F_\lambda(x,t)=-\frac{\lambda^2}{2}C_\Theta(x)+\frac{2ih\lambda L}{\pi}f(x,t),
\end{gather}
where 
\begin{gather}
C_\Theta(x)=\langle 0|\Theta(x,t)^2|0\rangle=-\frac{1}{2K}\ln\left(\frac{2\pi\alpha}{L}\sin\left(\frac{\pi x}{L}\right)\right)
\label{ctheta}
\end{gather}
is time independent. The other function $f(x,t)$ is found to be related to $g(x,t)$ from the particle density as
\begin{gather}
f(x,t)\equiv \int_0^L \frac{dx'}{2L}\langle 0|\left\{\Theta(x,t),\phi(x')\right\}|0\rangle=-\frac{g(vt,x/v)}{K},
\label{fx}
\end{gather}
which is independent from the LL parameter $K$, since the two bosonic fields get renormalized  in an opposite fashion in Eqs. \eqref{phix} and \eqref{thetax}.
This gives
\begin{gather}
j_0(x,t)=\frac{vhK}{\pi^2}\sum_{\sigma=\pm}\sigma \ln\left|\tan\left(\frac{\pi (x+\sigma vt)}{2L}\right)\right|,
\label{j0}
\end{gather}
which vanishes at $t=0$ as expected. It exhibits light cones similarly to the particle density.

\section{Switching on the non-hermitian term ($h=0\to h\ne 0 $)}\label{sec:on}

In the alternative protocol, we follow the reverse procedure, starting from the Hermitian ground state with no imaginary vector potential ($h=0$) and then abruptly switch on the non-Hermitian parameter $h$. This results in a true non-Hermitian quench, as the time evolution becomes non-unitary due to the presence of the imaginary vector potential. Here, the initial state is the bosonic vacuum $|0\rangle$, and the time evolution is dictated by Eq. \eqref{hamboson}, which can be expressed using the inverse similarity transformation as $SH_bS^{-1}$. This will be dubbed the 'switch on' protocol.
The time evolved wavefunction is 
\begin{gather}
|\Phi(t)=\frac{S\exp(-iH_bt)S^{-1}|0\rangle}{\sqrt{\langle 0|S^{-1}\exp(iH_bt)S^2\exp(-iH_bt)S^{-1}| 0\rangle}}.
\end{gather}
Using Eq. \eqref{identity} twice and the time evolution, we get
\begin{gather}
G_\lambda(x,t)=\nonumber\\
=\frac{\left\langle 0 \left|\exp\left( i\lambda \phi(x,t)+\frac{2h}{\pi} \int_0^L\left(\phi(x',t)-\phi(x')\right)dx'\right)\right |0 \right\rangle}
{\left\langle 0 \left|\exp\left(\frac{2h}{\pi} \int_0^L\left(\phi(x',t)-\phi(x')\right)dx'\right)\right |0 \right\rangle}.
\end{gather}
After taking the expectation value, this reads as
\begin{gather}
\ln G_\lambda(x,t)=-\frac{\lambda^2}{2}\langle 0|\phi(x,t)^2|0\rangle+
\nonumber\\
+\frac{ih\lambda L}{\pi} \int_0^L \frac{dx'}{L}\langle 0|\left\{\phi(x,t),\phi(x',t)-\phi(x')\right\}|0\rangle
\end{gather}
which differs from Eq. \eqref{g1} by the equal time autocorrelator  $\sim \int_0^L \langle 0|\{\phi(x,t),\phi(x',t)\}|0\rangle dx'$, which is independent of time.
Putting everything together, we
obtain
\begin{gather}
\ln G_\lambda(x,t)=-\frac{\lambda^2}{2}C_\phi(x)+\frac{2ih\lambda L}{\pi}\left(g(x,0)-g(x,t)\right).
\end{gather}

\begin{figure}[t!]
\centering
\includegraphics[width=8cm]{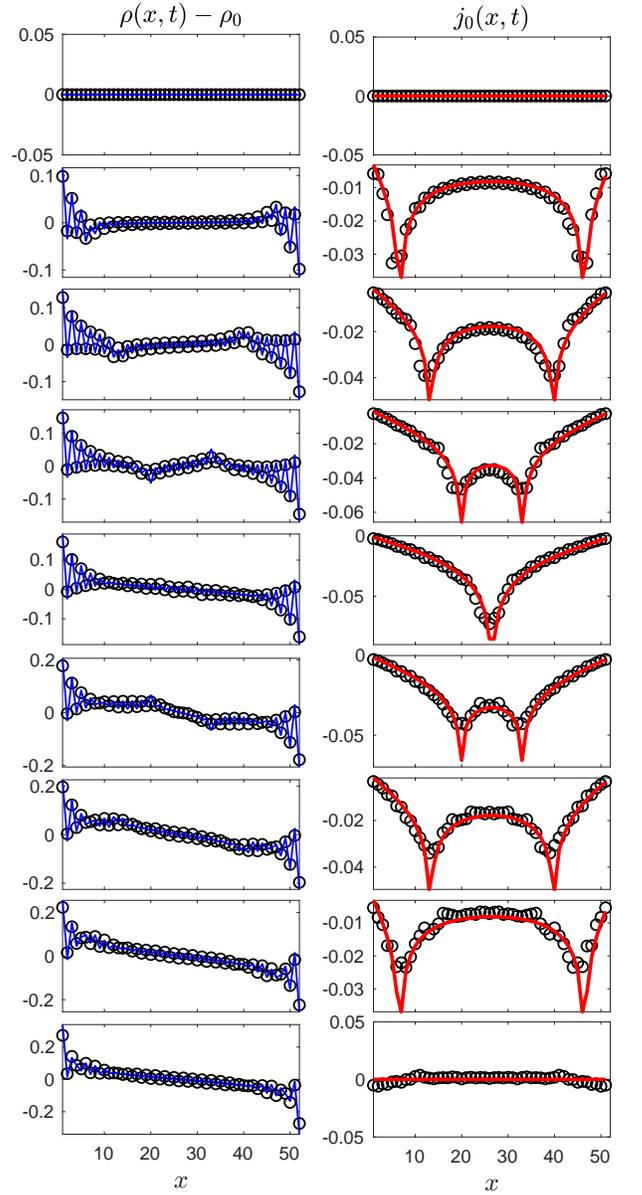}
\caption{Real space density profile (left) and current (right) for the non-interacting case with $U=0$, $L=52$ and times $t=k~L/8v$ with $k=0:1:8$ from top to bottom 
after switching on
$h=0.1$. The Friedel oscillations are plotted using $c=0.43$ and $\delta=0$. The circles denote the tight binding numerics, the only overall fitting parameter is $c$
for the oscillating part of the
particle density, the long wavelength parts contain no fitting parameter, as expected from Eqs. \eqref{n01} and \eqref{j01}.}
\label{friedel1}
\end{figure}

\begin{figure}[t!]
\centering 
\includegraphics[width=8cm]{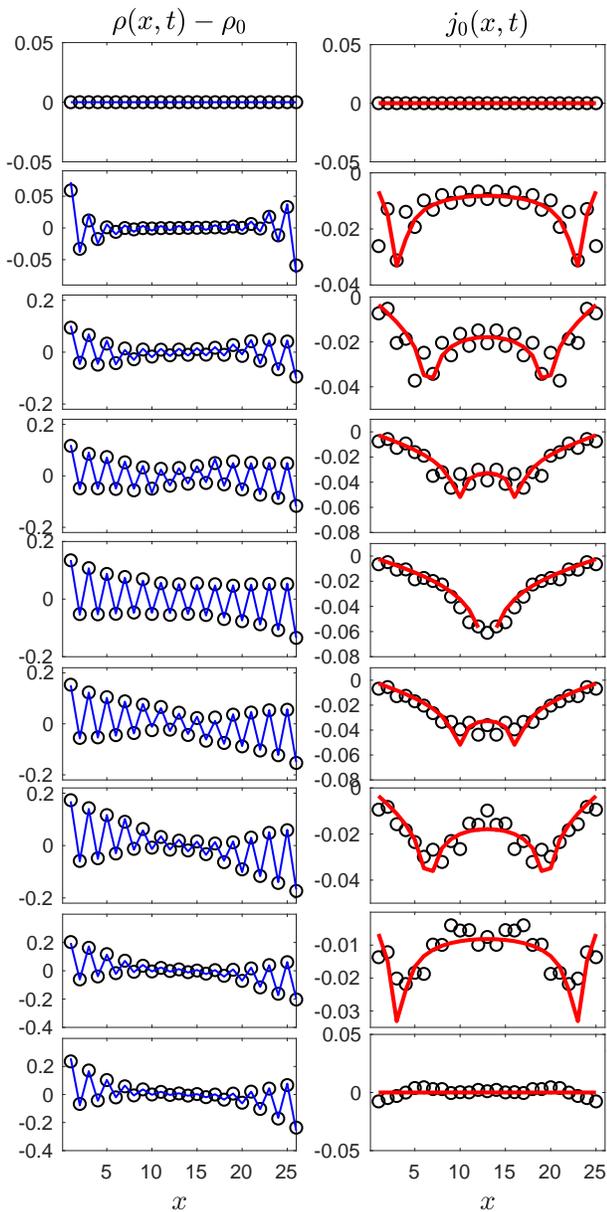}
\caption{Real space density profile (left) and current (right) for the interacting case with $U=0.5J$, $L=26$ and times $t=k~L/8v$ with $k=0:1:8$ from top to bottom
after switching on
$h=0.1$. The Friedel oscillations are plotted using $c=0.43$ and $\delta=0$. The circles denote the many-body ED results, the only overall fitting parameter is $c$
for the oscillating part of the
particle density, the long wavelength parts contain no fitting parameter in accord with Eqs. \eqref{n01} and \eqref{j01}.}
\label{friedel2}
\end{figure}

The homogeneous part of the particle density builds up as
\begin{gather}
n_0(x,t)=-\frac{Kh}{\pi^2}\left(2\ln\left(\tan\left(\frac{\pi x}{2L}\right)\right)-\right.\nonumber\\
-\left.\sum_{\sigma=\pm}\ln\left|\tan\left(\frac{\pi (x+\sigma vt)}{2L}\right)\right|\right).
\label{n01}
\end{gather}
Combining all elements, the overall time-dependent particle density is
\begin{gather}
\rho(x,t)=\rho_0+n_0(x,t)+c\left(\frac{\pi\alpha}{2L\sin\left(\frac{\pi x}{L}\right)}\right)^K\times\nonumber\\
\times\sin\left(2k_Fx+\frac{4hL}{\pi}(g(x,t)-g(x,0))\right),
\label{totaln1}
\end{gather}
where $\rho_0$ represents the homogeneous background. These are plotted in Figs. \ref{friedel1} and \ref{friedel2}.
These results indicate that at least for small $h$, the strong localization of eigenstates to one end of the chain through the non-hermitian skin effect does not appear, but rather
a ballistic propagation of light cones characterizes the dynamics.

We have also evaluated the current after switching on the non-hermitian term.
For the generating function, we obtain
\begin{gather}
F_\lambda(x,t)=\nonumber\\
=\frac{\left\langle 0 \left|\exp\left( i\lambda \Theta(x,t)+\frac{2h}{\pi} \int_0^L\left(\phi(x',t)-\phi(x')\right)dx'\right)\right |0 \right\rangle}
{\left\langle 0 \left|\exp\left(\frac{2h}{\pi} \int_0^L\left(\phi(x',t)-\phi(x')\right)dx'\right)\right |0 \right\rangle}.
\end{gather}  
After taking the expectation value, we get a term of the form $\langle 0|\{\Theta(x,t),\phi(x',t)-\phi(x')\}|0\rangle$. From this, the equal time anticommutator
vanishes identically in the ground state, i.e. $\langle 0|\{\Theta(x,t),\phi(x',t)\}|0\rangle=\langle 0|\{\Theta(x),\phi(x')\}|0\rangle=0$.
Then, we get
\begin{gather}
\ln F_\lambda(x,t)=-\frac{\lambda^2}{2}C_\Theta(x)-\frac{2ih\lambda L}{\pi}f(x,t),
\end{gather}
where $C_\Theta(x)$ and $f(x,t)$ are given by Eqs. \eqref{ctheta} and \eqref{fx}.
This yields
\begin{gather}
j_0(x,t)=-\frac{vhK}{\pi^2}\sum_{\sigma=\pm}\sigma \ln\left|\tan\left(\frac{\pi (x+\sigma vt)}{2L}\right)\right|,
\label{j01}
\end{gather}
which  is identical to the previous case (except for an overall minus sign) 
when the non-hermitian term is switched off. 
The ensuing density and current pattern is very similar to that in Fig. \ref{friedel} albeit the particle density picks up an additional time independent tilt from the 
first term on the r.h.s. of Eq.\eqref{n01}, 
while the current 
simply flows in the opposite direction compared to
Fig. \ref{friedel}.

\section{Continuity equation}\label{sec:ce}

The continuity equation states that the local density changes in time when local currents flow or some external source or sink is present~\cite{schomerus}.
By decomposing the non-hermitian Hamiltonian as $H=H_0+iV$ with both $H_0$ and $V$ hermitian,
the expectation value of the local density, $n$ is
\begin{gather}
\langle n\rangle=\frac{\langle \Phi(t)|n|\Phi(t)\rangle}{\langle \Phi(t)|\Phi(t)\rangle},
\label{expval}
\end{gather}
where $|\Phi(t)\rangle=e^{-i H t}|\Phi_0\rangle$.
The time derivative of this expectation value reads as~\cite{graefe2008}
\begin{gather}
\partial_t \langle n\rangle =i\langle \left[H,n\right]\rangle+\langle \left\{V,n\right\}\rangle-2\langle n \rangle\langle V\rangle,
\end{gather}
where $[A,B]$ stands for the commutator.
The first term on the right hand side represents the conventional term for hermitian
systems, the second term with the anticommutator stems from the non-hermitian contribution, namely from the
interaction with the environment, while the very last term originates from the explicit normalization of the wavefunction in Eq. \eqref{expval}. Then, the continuity equation is
\begin{gather}
\partial_t \langle n\rangle +\partial_x \langle j\rangle =\langle \left\{V,n\right\}\rangle -2\langle n \rangle\langle V\rangle.
\label{ce}
\end{gather}
For the 'switch off' protocol, the time evolution is dictated by a hermitian Hamiltonian, thus $V=0$ and the continuity equation holds naturally, as expected.
For the 'switch on' procedure, on the other hand, $V=-vhK \int_0^L \Pi(x)dx=vhK (\Theta(0)-\Theta(L))/\pi$ from Eq. \eqref{hamboson}.
Using this and the long wavelength density operator $n=\partial_x \phi(x)/\pi$, the r.h.s. of Eq. \eqref{ce} indeed vanishes, in accordance with Eq. \eqref{n01} and
\eqref{j01}, which make the l.h.s. of Eq. \eqref{ce} vanish.

\section{Numerics}\label{sec:numerics}

The Hatano-Nelson model~\cite{hatanonelson2,hatanonelson1} consists of fermions hopping in one dimension in the presence of an imaginary vector potential. The 
interacting
many-body version of the Hamiltonian is
\begin{gather}
H_{HN}=\sum_{n=1}^{N-1} \frac J2 \exp(ah)c^\dagger_nc_{n+1}+\frac J2 \exp(-ah) c^\dagger_{n+1}c_n+\nonumber \\
+U c^\dagger_nc_{n}c^\dagger_{n+1}c_{n+1},
\label{hamiltontb}
\end{gather}

where $J$ is the uniform hopping, $h$ is the constant imaginary vector potential and $a$ represents the lattice constant,
$N$ is the total number of lattice sites and we consider
open boundary condition (OBC), $U$ represents the the nearest-neighbour interaction between particles.
The first term describes the hopping of particles from site $n$ to site $n+1$, while the second term describes the opposite hopping direction. 
We consider half filling with $N/2$ fermions populating the lattice.
 The model is PT-symmetric\cite{bender2007} and possesses a real spectrum for OBC, and the minimal energy
configuration is the ground state with many-body wavefunction $|\Psi\rangle$.
In the presence of finite $U$, the LL parameter is $K=\pi/2/(\pi-\arccos(U/J))$ while $v$ can be obtained from the $2L/v$ time periodicity of
the density and current patterns. More precisely, the current vanishes identically for the first time after the switch on or off at $t=L/v$.
We assume that the above value of $K$ remains valid also for small $h$ as well.

We study $H_{HN}$ numerically by solving the time-dependent Schr\"odinger equation for $U=0$.
The initial many-body (i.e.~$N/2$-body) state $\Psi_0$ is a Slater determinant made from the 
single particle eigenstates of Eq.~\eqref{hamiltontb} as $\phi_n$ with $h\neq 0$ but $\langle \phi_n|\phi_n'\rangle\neq \delta_{n,n'}$ due to non-hermiticity.
Then, the 'switch off' protocol is followed at the single particle level as $\phi_n(t)=\exp[-iH_{HN}t]\phi_n$ for $t>0$ with $h=0$ in $H_{HN}$.
The 'switch on' protocol is slightly different: the initial wavefunction are orthogonal due to the $h=0$ hermitian initial Hamiltonian as $\langle 
\phi_n|\phi_n'\rangle=\delta_{n,n'}$. They become non-orthogonal only due to the non-unitary time evolution from $h\neq 0$.
The corresponding results are shown in Fig. \ref{friedel1}.

The time-evolved many-body wavefunction, $\Psi(t)$ remains a Slater determinant built up from these 
time-dependent single particle functions, which are not orthogonal for both protocols.
After time $t$, we evaluate numerically the change in the density profile~\cite{carmichael,daley,ashidareview}
as
\begin{gather}
\rho(n,t)=\frac{\langle\Psi(t)|c^\dagger_nc_n|\Psi(t)\rangle}{\langle\Psi(t)|\Psi(t)\rangle},
\label{eq:rho}
\end{gather}
where the denominator is required as it accounts for the non-unit norm of the many-body wavefunction~\cite{graefe2008} and the homogeneous background density is $\rho_0=1/2$.
For the 'switch-off' protocol, this differs from unity but does not change in time while for the 'switch on' protocol, it would start from unity and change with time.
Since the many-body wavefunction is a Slater determinant, $c_n$ in the numerator acts separately on the single particle wavefunctions.
However, due to the non-orthogonality of $\phi_n(t)$, the overlap of the other wavefunctions,
not acted on by $c_n$, has to be evaluated as well and can give non-trivial (i.e. not 0 or 1) contribution.
We also evaluate in a similar fashion the time evolved local particle current operator from
\begin{gather}
j_n=iJ\left(c^\dagger_{n+1}c_n -c^\dagger_{n}c_{n+1}\right)/2.
\end{gather}

When dealing with finite U, we use many-body exact diagonalization (ED) on small systems (Fig. \ref{friedel2}). We also utilize the DMRG algorithm~\cite{White-1992} to search for the 
many body ground state within the matrix product states (MPS) framework. Subsequently, 
the MPS wave function $\Psi(t)$ is time-evolved, and the density and current profiles are calculated, as described in Eq.~\eqref{eq:rho}, see Fig. \ref{friedel1} as well. In the 
'switch off' approach, the time 
evolution is unitary, and the wave function maintains its normalization at any later time. Conversely, in the 'switch on' protocol, the wave function is no longer normalized due to the non-unitary evolution.

\section{Conclusions}
Our research focused on examining the behavior of the many-body interacting Hatano-Nelson model with open boundary condition following a quantum quench. 
Using abelian bosonization, we derived  analytical expressions for the spatio-temporal profiles of both density and current along the chain. Two distinct quench protocols were considered, 
one with a unitary evolution after switching off the non-hermitian term and one with a non-unitary evolution after switching on the imaginary vector potential. 
Our findings revealed that in both cases, the dynamics exhibited a ballistic behavior of light cone propagation, starting from the ends of the chain. This influenced the homogeneous particle density, the Friedel oscillations as well as the particle current.
The continuity equation involving the the long wavelength part of the density and current remains satisfied, in spite of non-hermiticity\cite{schomerus}.
Interestingly, we found that the magnitude of the current is the same for both protocols. 
Our results were supported by numerical methods, such as exact diagonalization or time evolving block decimation.

\begin{acknowledgments}
This research is supported by the National Research, Development and 
Innovation Office - NKFIH  within the Quantum Technology National Excellence 
Program (Project No.~2017-1.2.1-NKP-2017-00001), K134437, K142179 by the BME-Nanotechnology 
FIKP grant (BME FIKP-NAT), and by a grant of the Ministry of Research, Innovation and
 Digitization, CNCS/CCCDI-UEFISCDI, under projects number PN-III-P4-ID-PCE-2020-0277 and 
 under the project for funding the excellence, Contract No. 29 PFE/30.12.2021.
M.A.W has also been supported by the Janos Bolyai
Research Scholarship of the Hungarian Academy of Sciences
and by the \'UNKP-22-5-BME-330 New National
Excellence Program of the Ministry for
Culture and Innovation from the source of the National
Research, Development and Innovation Fund. 
\end{acknowledgments}

\bibliographystyle{apsrev}
\bibliography{wboson1}

\end{document}